\documentclass[10pt,twoside,BCOR7mm,DIV15,headinclude,footexclude,cleardoubleempty,idxtotoc]{scrartcl}

\usepackage[english]{babel}
\usepackage{graphicx}
\usepackage{hyperref}
\usepackage{scrpage2}
\usepackage{hyperref}
\usepackage{ifthen}

\makeatletter
\renewcommand{\@biblabel}[1]{}
\renewcommand{\@cite}[2]{%
{#1\ifthenelse{\boolean{@tempswa}}{,#2}{}}}
\makeatother

\hypersetup{breaklinks=true
,colorlinks=true,linkcolor=black,urlcolor=blue
,citecolor=black}

\ofoot{\thepage}
\ifoot{}

\setheadsepline{1pt}

\setkomafont{pagehead}{\normalfont\sffamily}
\setkomafont{pagenumber}{\normalfont\rmfamily}

\usepackage{booktabs}
\usepackage{amsmath}
\usepackage{amssymb}
\usepackage{multicol}
\usepackage{float}

\makeatletter
\newcommand{\listofcontributions}{\@starttoc{con}}

\newcommand{\l@contribution} {\@dottedtocline{1}{1.5em}{2.3em}}
\makeatother

\newenvironment{contribution}{
\setcounter{section}{0}
\setcounter{figure}{0}
\setcounter{table}{0}
\begin{flushleft}
{\em Clumping in Hot Star Winds \\
W.-R.\ Hamann, A.\ Feldmeier \& L.\ Oskinova, eds.\\
Potsdam: Univ.-Verl., 2007 \\
URN: http://nbn-resolving.de/urn:nbn:de:kobv:517-opus-13981
} 
\end{flushleft}
}{
\newpage
\lehead{}
\rohead{}
}

\begin{document}

\setlength{\baselineskip}{2.5ex}

\begin{contribution}


\lehead{D.\ H.\ Cohen, M.\ A.\ Leutenegger, \& R.\ H.\ D.\ Townsend}


\rohead{Resolved X-ray emission line profiles}

\begin{center}


{\LARGE \bf Quantitative analysis of resolved X-ray emission line profiles of O stars
}\\

\medskip


{\it\bf D.\ H.\ Cohen$^1$, M.\ A.\ Leutenegger$^2$, \& R.\ H.\ D.\ Townsend$^3$}\\


{\it $^1$Department of Physics and Astronomy, Swarthmore College, Swarthmore, Pennsylvania, United States}\\
{\it $^2$Columbia Astrophysics Laboratory, Columbia University, New York, New York, United States}\\
{\it $^3$Bartol Research Institute, University of Delaware, Newark, Delaware, United States}

\begin{abstract}
  By quantitatively fitting simple emission line profile models that
  include both atomic opacity and porosity to the {\it Chandra} X-ray
  spectrum of $\zeta$ Pup, we are able to explore the trade-offs
  between reduced mass-loss rates and wind porosity.  We find that
  reducing the mass-loss rate of $\zeta$ Pup by roughly a factor of
  four, to $1.5 \times 10^{-6}~\mathrm{M}_{\odot}~\mathrm{yr}^{-1}$,
  enables simple non-porous wind models to provide good fits to the
  data.  If, on the other hand, we take the literature mass-loss rate
  of $6 \times 10^{-6}~\mathrm{M}_{\odot}~\mathrm{yr}^{-1}$, then to
  produce X-ray line profiles that fit the data, extreme porosity
  lengths -- of $h_{\infty} \approx 3~\mathrm{R}_{\ast}$ -- are
  required.  Moreover, these porous models do {\it not} provide better
  fits to the data than the non-porous, low optical depth models.
  Additionally, such huge porosity lengths do not seem realistic in
  light of 2-D numerical simulations of the wind instability.
\end{abstract}
\end{center}

\begin{multicols}{2}

\section{Introduction}

The X-ray emission lines from O stars are wind-broadened but are
surprisingly symmetric.  Asymmetry should arise due to the
preferential absorption of red-shifted photons, which are produced on
the far side of the wind and seen through a larger column density of
cold, absorbing material. The very modest asymmetry in the
observations implies that mass-loss rates are lower than has been
presumed, which is, of course, in line with other recent observations.
Alternatively, it has been suggested that the mass-loss rates are
actually high, but the effective opacity of the wind is reduced due to
porosity, or macroclumping (Oskinova et al.\ \cite{ofh2006}).

In this very brief paper, in which we focus on one strong
representative line in the {\it Chandra} grating spectrum of $\zeta$
Pup, we explore whether the measured profile shapes can discriminate
between mass-loss reduction and porosity.  Even if they cannot, by
fitting models to data we can quantitatively explore the trade-off
between the key parameters: fiducial optical depth, $\tau_{\ast}
\equiv \kappa\dot{M}/4{\pi}v_{\infty}R_{\mathrm \ast}$, and the
terminal porosity length, $h_{\rm \infty}$.  Here we do this by
fitting the Owocki \& Cohen (\cite{oc2001}) profile model to one
strong line in the {\it Chandra} HETGS/MEG spectrum of $\zeta$ Pup. We
then fit a modified profile model, where the opacity is adjusted for
porosity according to Owocki \& Cohen (\cite{oc2006}).

\section{Fitting Models to the Data}

There are 560 counts in the Fe XVII line shown in Fig.\
\ref{fig:cohen_smoothfit}, accumulated during an exposure time of 68
ks. The line is well resolved and clearly asymmetric, with the
characteristic blue shifted and skewed line shape that is expected
from a spherically symmetric, beta-velocity wind with embedded X-ray
emission and continuum absorption from the cool, dominant component of
the wind.  The first model we fit to the data is the simplest -- a
non-porous Owocki \& Cohen (\cite{oc2001}) model where the optical
depth depends on the atomic opacity. We note that this model does not
preclude microclumping of the sort that affects density-squared
emission diagnostics, but does not affect column-density based
diagnostics like X-ray emission line profiles.  However, this standard
profile model does {\it not} include large scale porosity, or
macroclumping, with $h \equiv {\ell}/f > r$, where $h$ is the porosity
length, defined as the clump size scale to the volume filling factor
(Owocki \& Cohen \cite{oc2006}). We do fit porous models later in this
paper.

We fit the model, along with a fixed power-law continuum, within
XSPEC, to the line over the wavelength range shown in Fig.\
\ref{fig:cohen_smoothfit}.  The fitting routine in XSPEC adjusts the
free parameters of the model: $R_{\mathrm 0}$, the radius below which
there is no X-ray emission, $\tau_{\ast}$, the fiducial wind optical
depth, and the normalization, until the global minimum of the fit
statistic -- here the C statistic (Cash \cite{Cash1979}), which is the
maximum likelihood statistic for Poisson-distributed data -- is found.
The best-fit model, with $R_{\mathrm 0} = 1.5 \pm .2 ~{\mathrm
  R_{\ast}}$ and $\tau_{\ast} = 2.0 \pm .5$, is plotted over the data
in Fig.\ \ref{fig:cohen_smoothfit}.

The uncertainties on the parameters can be estimated by evaluating
${\Delta}C$, the difference between the C statistic value for a given
choice of parameters and that of the best-fit model.  The distribution
of ${\Delta}C$ is the same as that of ${\Delta}{\chi}^2$ and a
specific value corresponds to a formal confidence limit.  By drawing
contours of constant ${\Delta}C$ in the parameter space of interest,
we can see the extent of a given parameter's possible range and the
correlation between parameters, as we show in Fig.\
\ref{fig:cohen_smoothcontour}.

\begin{figure}[H]
\begin{center}
\includegraphics[width=\columnwidth]{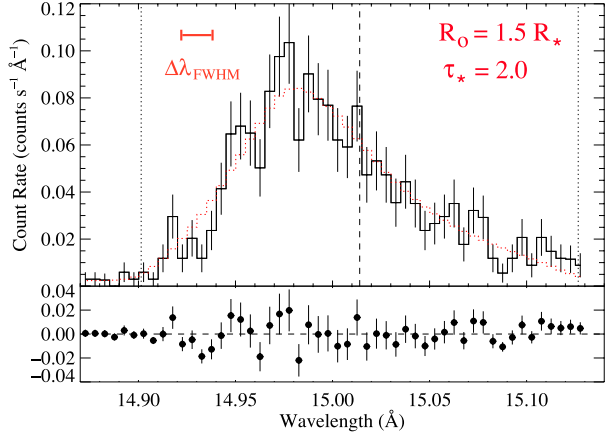}
\caption{An Fe XVII emission line, formed in a thermal plasma at a few $10^6$ K, with Poisson error bars.  The approximate spectral resolution is indicated by the orange bar. The best-fit non-porous model is shown. 
  \label{fig:cohen_smoothfit}}
\end{center}
\end{figure}

\begin{figure}[H]
\begin{center}
\includegraphics[width=\columnwidth]{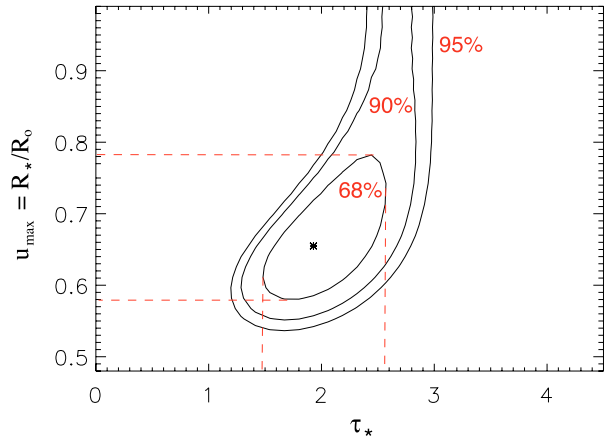}
\caption{These confidence contours in the $R_{\mathrm 0}$ - $\tau_{\ast}$
  parameter space of the fit of the non-porous model to the Fe XVII
  line, shown in Fig.\ \ref{fig:cohen_smoothfit}, enclose 68\%, 90\%,
  and 95\% of the model parameter space where the true model
  parameters lie, given the data and the assumed model. 
  \label{fig:cohen_smoothcontour}}
\end{center}
\end{figure}

The value of $\tau_{\ast}$ we derive from the model fitting,
$\tau_{\ast} = 2.0 \pm .5$, implies a mass-loss rate of $\dot{M} = 1.5
\pm .4 \times 10^{-6}~\mathrm{M}_{\odot}~\mathrm{yr}^{-1}$, given an
X-ray opacity at 15 \AA\/ of 70 cm$^2$ g$^{-1}$ (Waldron et al.\
\cite{Waldron1998}). If, on the other hand, we take the mass-loss rate
of $\zeta$ Pup to be $\dot{M} = 6 \times
10^{-6}~\mathrm{M}_{\odot}~\mathrm{yr}^{-1}$ and use $R_{\ast} = 19~
{\mathrm R_{\odot}}$ and $v_{\infty} = 2250$ km s$^{-1}$ (Puls et al.\
\cite{Puls1996}), then we expect the wind to have a fiducial optical
depth of $\tau_{\ast} = 8$.  The best-fit model is favored over a
$\tau_{\ast} = 8$ model with $> 99.99$\% confidence (${\Delta}C =
79$).  In Fig.\ \ref{fig:cohen_2smoothmodels} we show the optimal
model with a fixed $\tau_{\ast} = 8$ ($R_{\mathrm 0}$ and
normalization are free parameters) along with the global best-fit
model.

\begin{figure}[H]
\begin{center}
\includegraphics[width=\columnwidth]{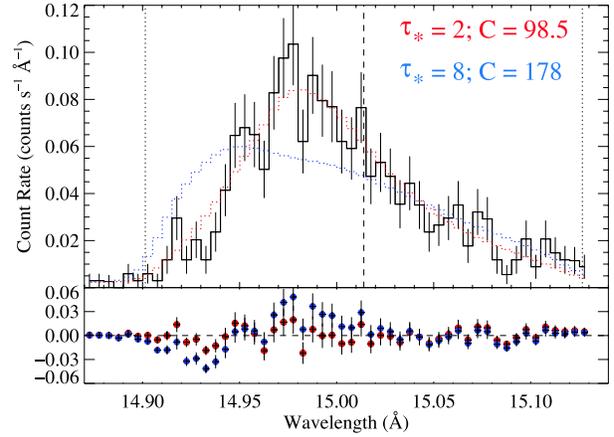}
\caption{The global best-fit non-porous wind model is shown in red
  and the best-fit model with the wind opacity expected based on the
  standard mass-loss rate is shown in
  blue.\label{fig:cohen_2smoothmodels}}
\end{center}
\end{figure}

\begin{figure}[H]
\begin{center}
\includegraphics[width=\columnwidth]{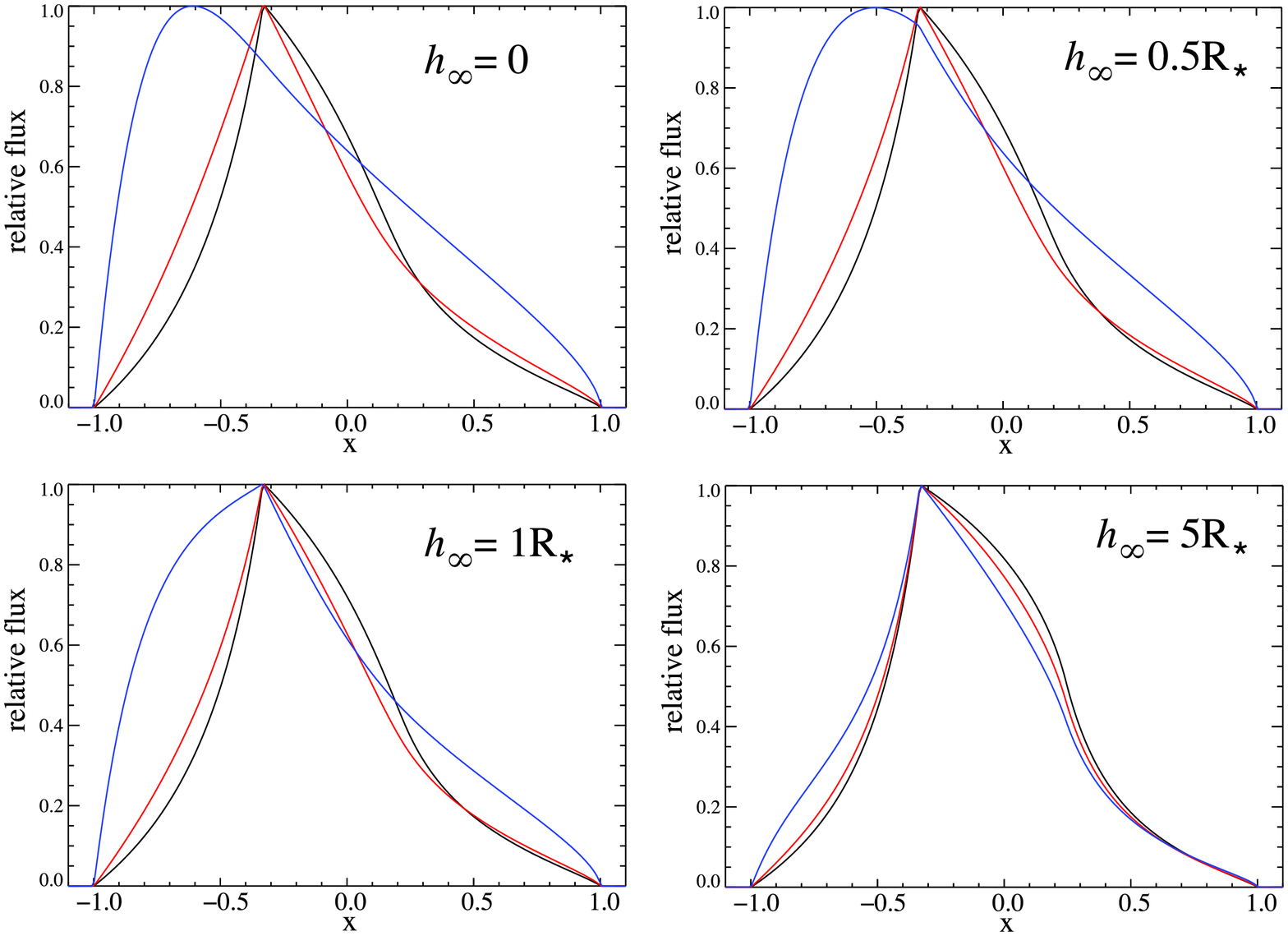}
\caption{Suite of porous line profile models, shown at infinite
  resolution.  Within each panel, three optical depth values are
  represented, $\tau_{\ast}=1,2,8$: black, red, blue. 
  \label{fig:cohen_4panel}}
\end{center}
\end{figure}

We next fit models that allow for porosity, but are otherwise
identical to the Owocki \& Cohen (\cite{oc2001}) models fit in the
previous subsection. In these porous models, the opacity is not
atomic, but rather geometric, and due to optically thick clumps of
size $\ell$ and filling factor $f$.  The optical depth is thus
modified according to ${\kappa}_{eff} = {\kappa}(1 -
e^{-\tau_c})/\tau_c$, where $\kappa$ is the microscopic opacity and
$\tau_c$ is the clump optical depth, $\tau_c =
\kappa\ell\langle\rho\rangle/f$, where $\langle\rho\rangle$ is the
local mean wind density (Owocki \& Cohen \cite{oc2006}). We
parameterize the porosity length as $h(r) =
h_{\infty}(1-R_{\ast}/r)^{\beta}$ and fix $\beta=1$, leaving the
terminal porosity length, $h_{\infty}$ as the single additional
parameter of the model.  As can be seen in Fig.\
\ref{fig:cohen_4panel}, porosity has only a weak effect, even at
$h_{\infty}=1$. Values above unity are required before the profiles
are strongly affected.

Now, by fitting this porous line profile model to the data, we can
examine the joint constraints on the key parameters: $\tau_{\ast}$ and
$h_{\infty}$, while allowing $R_{\mathrm 0}$ and the normalization to
also be free parameters of the model fit. When we do this, we find a
best-fit terminal porosity length of $h_{\infty} = 0.0$.  In other
words, the non-porous model is preferred over the porous one. In Fig.\
\ref{fig:cohen_porouscontour} we show the joint confidence limits on
$h_{\infty}$ and $\tau_{\ast}$.

\begin{figure}[H]
\begin{center}
\includegraphics[width=\columnwidth]{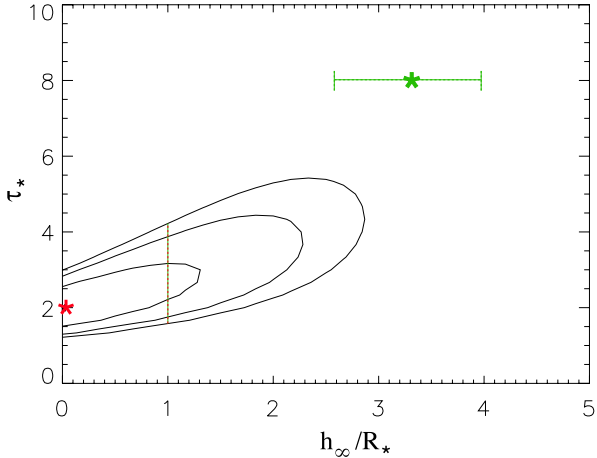}
\caption{The 68\%, 90\%, and 95\% confidence contours in $\tau_{\ast}$
  - $h_{\infty}$ parameter space, with the best-fit global model shown
  as the red star.  The green star is the location of the optimized
  model with $\tau_{\ast} = 8$ fixed at the value implied by the
  literature mass-loss rate. The green line denotes the 68\%
  confidence range on the terminal porosity length, $h_{\infty}$ when
  $\tau_{\ast} = 8$.  The vertical red line emphasizes that even with
  $h_{\infty} = 1$, the wind optical depth is only a little higher
  than the non-porous value of $\tau_{\ast} = 2.0$.
  \label{fig:cohen_porouscontour}}
\end{center}
\end{figure}

From this figure, we can see that the model with low optical depth and
no porosity is preferred with a high degree of formal significance
over the model with high optical depth and a large porosity length.
We can also see, from the shape of the confidence contours, that the
porosity length does not appreciably change the optical depth derived
from the data until $h_{\infty} > {\mathrm R_{\ast}}$.  This is in
accord with the expectations from the sensitivity of the model
profiles to the value of this parameter (as shown in Fig.\
\ref{fig:cohen_4panel}).


We stress that the porous model with the higher mass-loss rate
provides an adequate fit to the data, just not as good a fit as the
non-porous model.  
It is possible that a porous model with a different description of
porosity could fit the data as well as the non-porous model fit we
show.  However, we can definitively say two things from the model
fitting we have reported on here: (1) Porosity cannot explain the only
modestly asymmetric line profiles in $\zeta$ Pup any better than a
non-porous model with a reduced mass-loss rate can; if anything, it
provides a worse fit.  And, (2) in order to fit the data with a high
mass-loss rate, high porosity model, terminal porosity lengths of
several $\mathrm R_{\ast}$ are required.

Regarding this last point, two-dimensional simulations of the
line-driven instability show numerous very small structures, as the
compressed shells created by the instability break up laterally.
Constraints on moving emission bumps seen in WR winds using these 2-D
simulations shows that the clumps subtend about 3 degrees, or 1/20
steradian, implying linear clump scales of 1/20 the local stellar
radius, and thus quite small porosity lengths (Dessart \& Owocki
\cite{do2003,do2005}).


\end{multicols}

\end{contribution}


\end{document}